\documentclass[slac_one]{revtex4}
\usepackage{graphicx}
\usepackage{fancyhdr}
\newcommand{\om}{\ifmmode {w} \else {$w$}\fi}
\input babarsym

\begin{document}


\title{Measurements of Exclusive {\boldmath $B \to X_c\ell\bar{\nu}_{\ell}$} Decays and {\boldmath $|V_{cb}|$} at \babar\ } 

\author{David Lopes Pegna  (on behalf of the \babar\ collaboration)}
\affiliation{Princeton University, Princeton, NJ 08544}

\begin{abstract}
We present recent results on exclusive $\overline{B} \to X_c \ell^- \bar{\nu}_{\ell}$ decays and measurements of the CKM matrix element $|V_{cb}|$ based on data collected at the $\Upsilon(4S)$ resonance with the \babar\
detector at the \pep2\ $e^+e^-$ storage rings. 
\end{abstract}

\maketitle

\thispagestyle{fancy}


\section{Introduction} 

The study of $\overline{B} \to X_c \ell^- \bar{\nu}_{\ell}$ decays ($\ell$ = e or $\mu$ and charge conjugate modes are implied) is aimed at  understanding the dynamics of $b$-quark
semileptonic decays and to determine the relevant Cabibbo-Kobayashi-Maskawa matrix elements~\cite{CKM}.
The coupling strength of the weak $b \to c$ transition is proportional to $|V_{cb}|$, which 
has been measured in both inclusive semileptonic $B$ decays~\cite{incl} and in the exclusive transitions $\Bbar \to D\ell^-\bar{\nu}_{\ell}$ and $\Bbar \to D^*\ell^-\bar{\nu}_{\ell}$~\cite{excl}.
 The inclusive and exclusive determinations of $|V_{cb}|$ rely on different theoretical calculations. The former employs a parton-level calculation of the decay rate as a function of the strong coupling constant $\alpha_S$ and 
inverse powers of  the $b$-quark mass $m_b$. The latter relies on a parameterization of the decay form factors using 
Heavy Quark Symmetry and a non-perturbative calculation of the form factor normalization at the zero recoil 
(maximum momentum transfer squared). While the theoretical uncertainties in these two approaches are independent, the inclusive and exclusive experimental measurements make use of different techniques and have, to a large extent, uncorrelated statistical and systematic uncertainties. This independence makes the comparison of 
$|V_{cb}|$ from inclusive and exclusive decays a powerful test of our understanding of semileptonic decays. The latest 
determinations~\cite{pdg} differ by more than two standard deviations, and the inclusive determination is currently 
more than twice as precise as the exclusive one. 
Thus, improvements in the measurements of exclusive decays will strengthen this test. This is particularly true for the $\Bbar \to D\ell^-\bar{\nu}_{\ell}$ decay, where the experimental uncertainties dominate the determination of $|V_{cb}|$.
Based on current measurements~\cite{pdg} the rate of inclusive semileptonic $B$ decays exceeds the sum of the measured exclusive decay rates. While $\Bbar \rightarrow D \ell^- \bar{\nu}_{\ell}$ and  $D^* \ell^- \bar{\nu}_{\ell}$ decays account for about 70\% of this total, the contribution of decays to other charm states, including resonant and
 non-resonant $D^{(*)}\pi \ell^- \bar{\nu}_{\ell}$ (denoted by $D^{**} \ell^- \bar{\nu}_{\ell}$), is not yet well measured and may help to explain the inclusive-exclusive discrepancy. 
Improved measurements of $\Bbar \to X_c \ell^- \bar{\nu}_{\ell}$ decays will also benefit the accuracy of the extraction of $|V_{ub}|$, since analyses are extending into kinematic regions in which these decays represent a sizable background.

\section{Global Fit to {\boldmath $\Bbar \to D X \ell^-\bar{\nu}_{\ell}$} Decays}

\begin{figure}
\begin{minipage}[t]{0.43\linewidth} 
\centering 
\includegraphics[width=37mm]{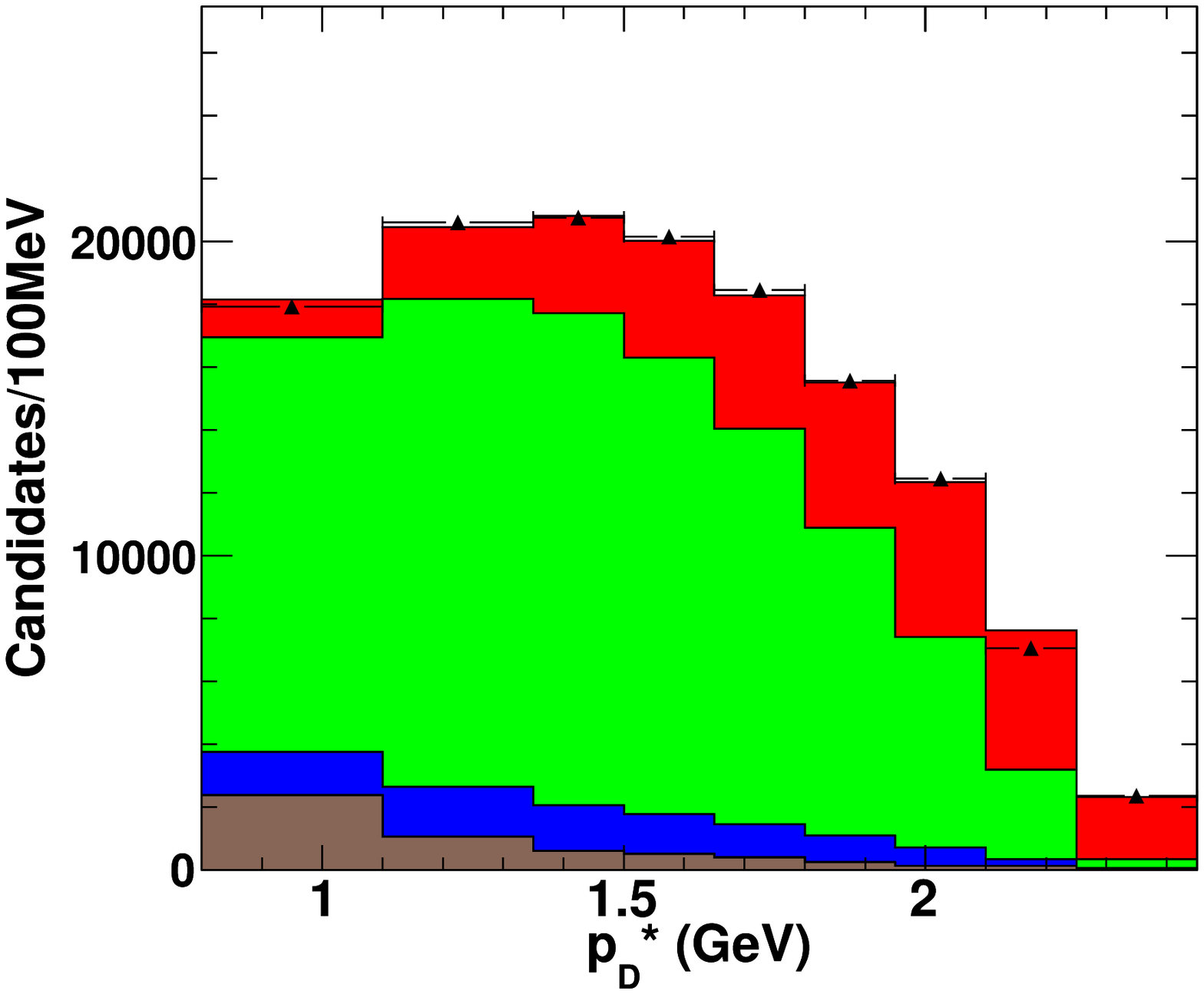}
\includegraphics[width=37mm]{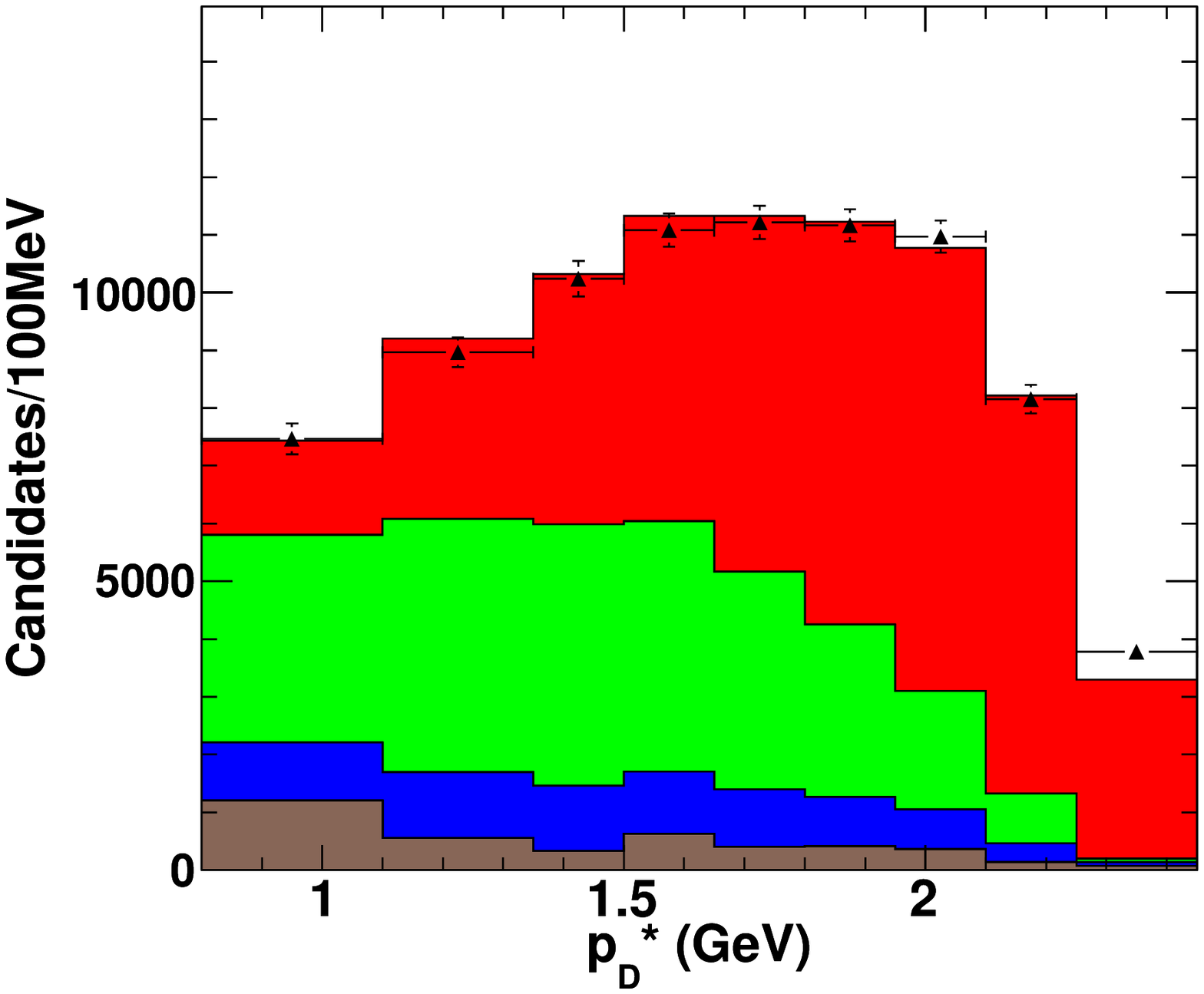}
\\
\includegraphics[width=37mm]{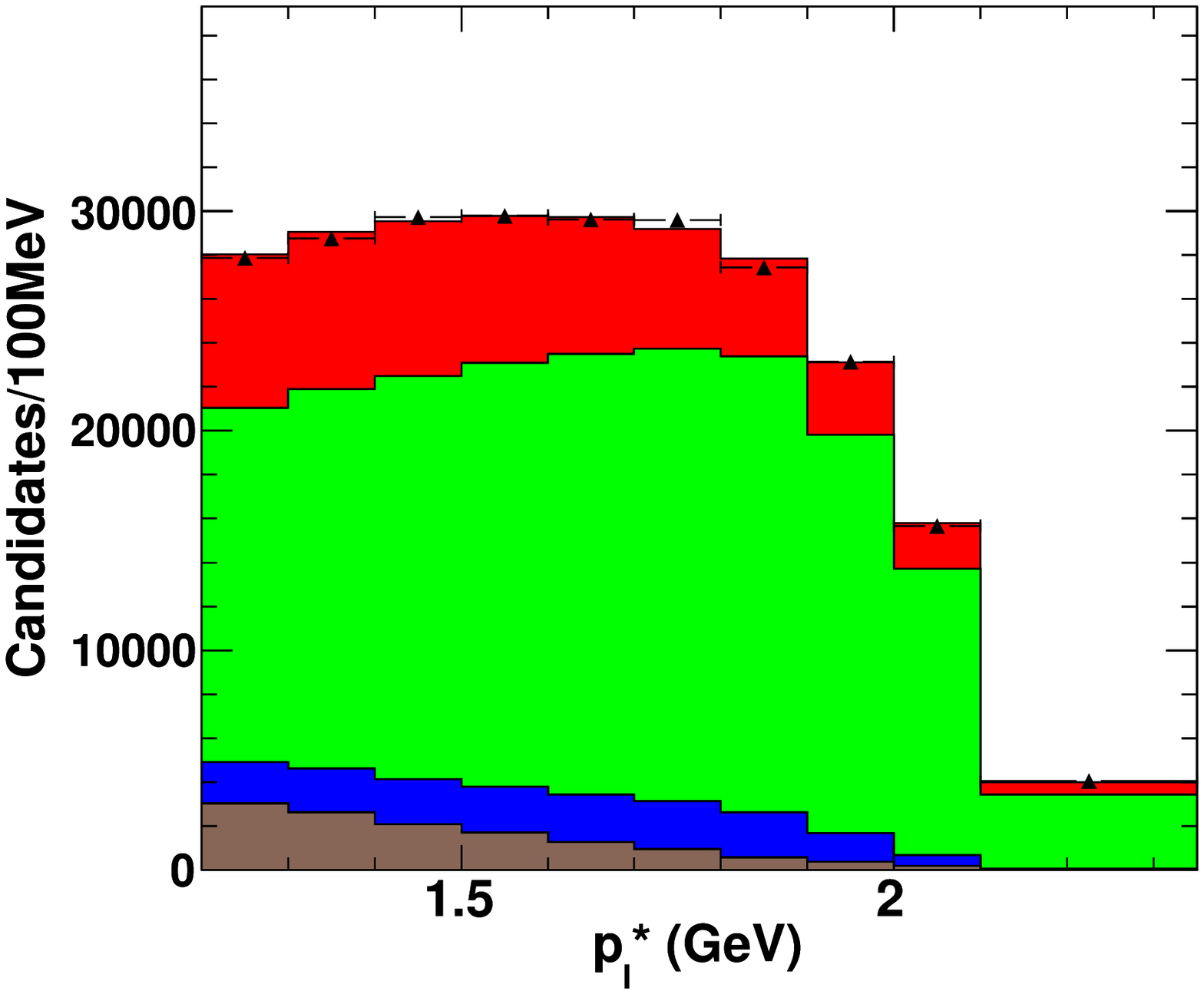}
\includegraphics[width=37mm]{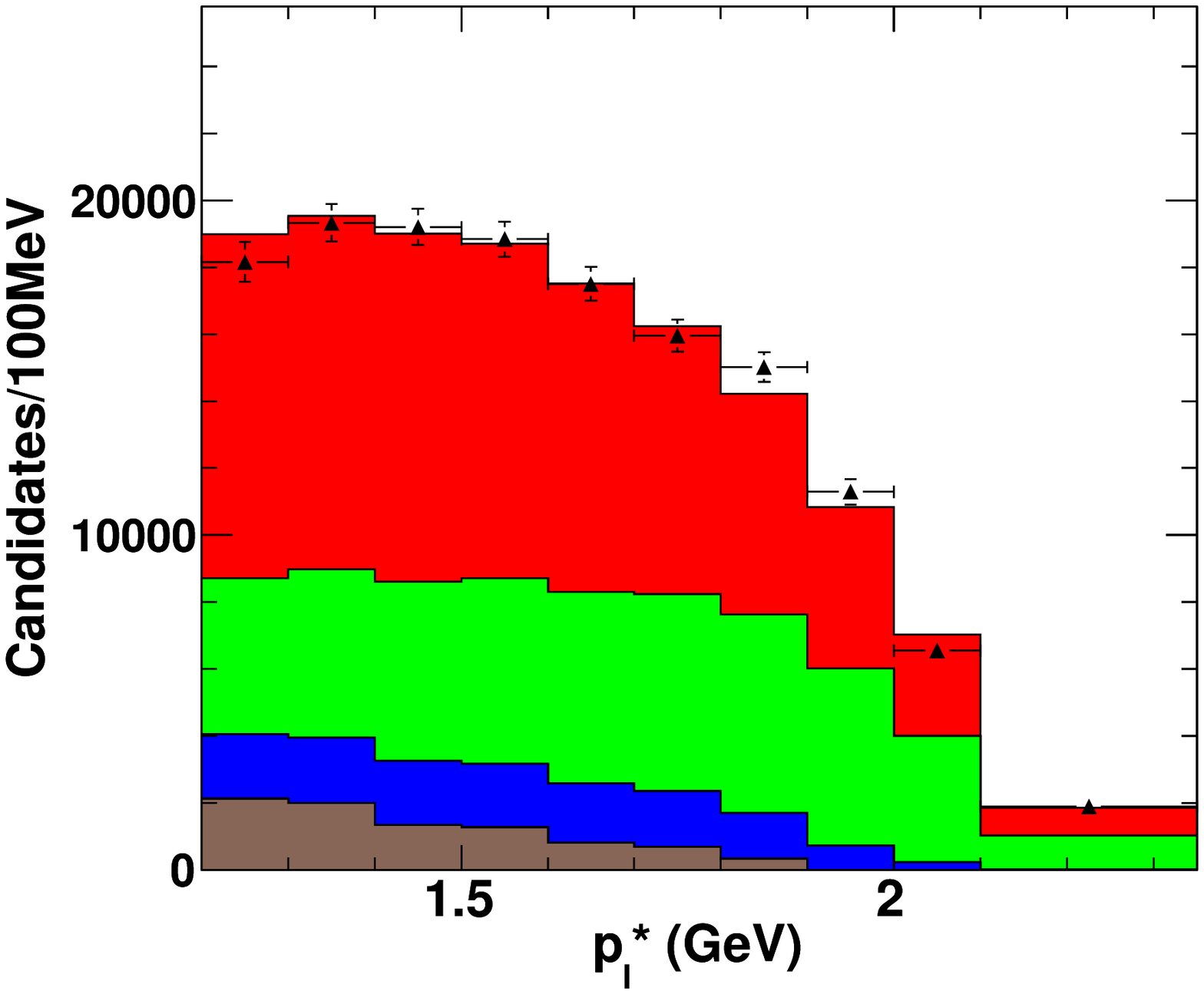}
\\
\includegraphics[width=37mm]{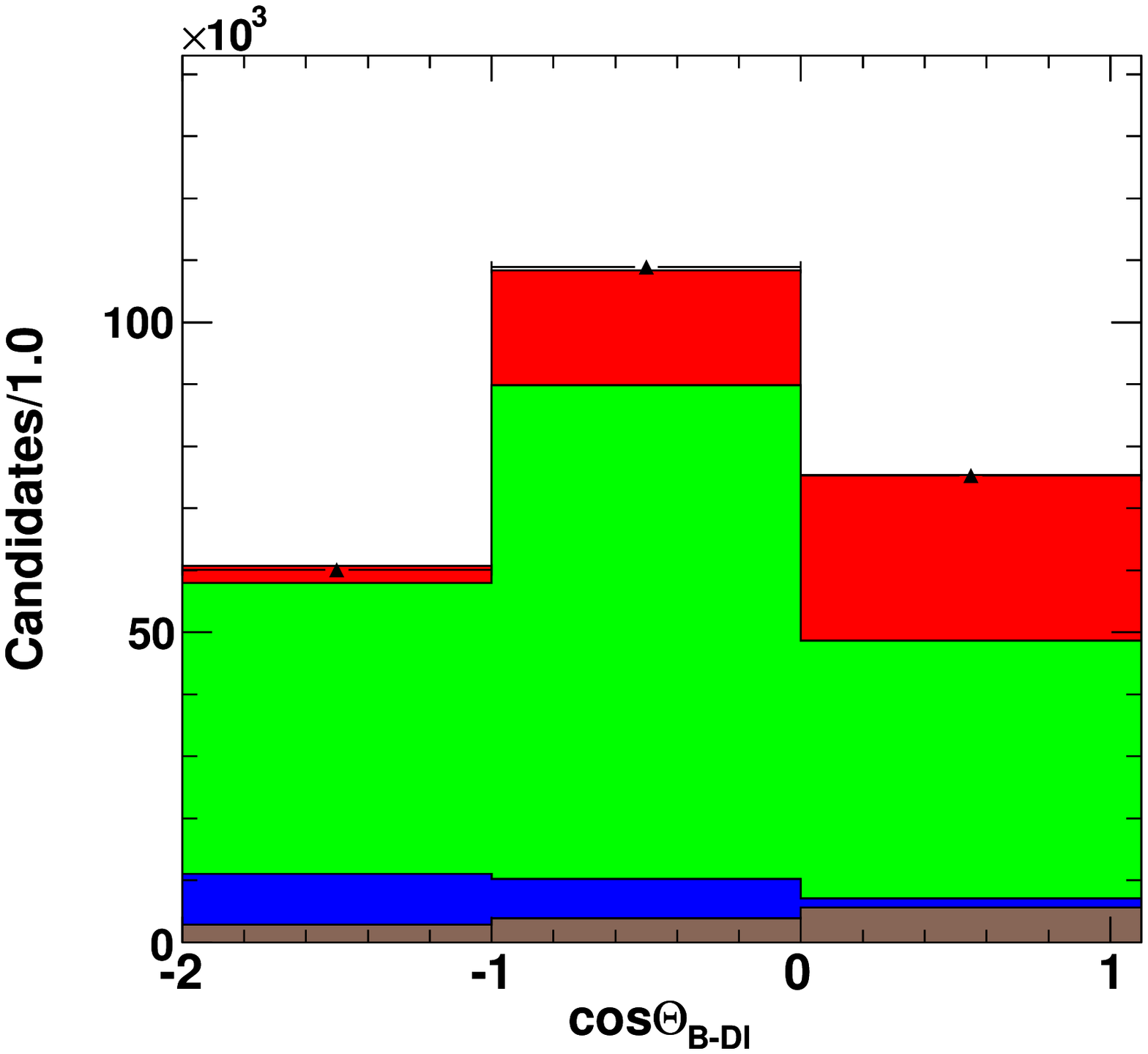}
\includegraphics[width=37mm]{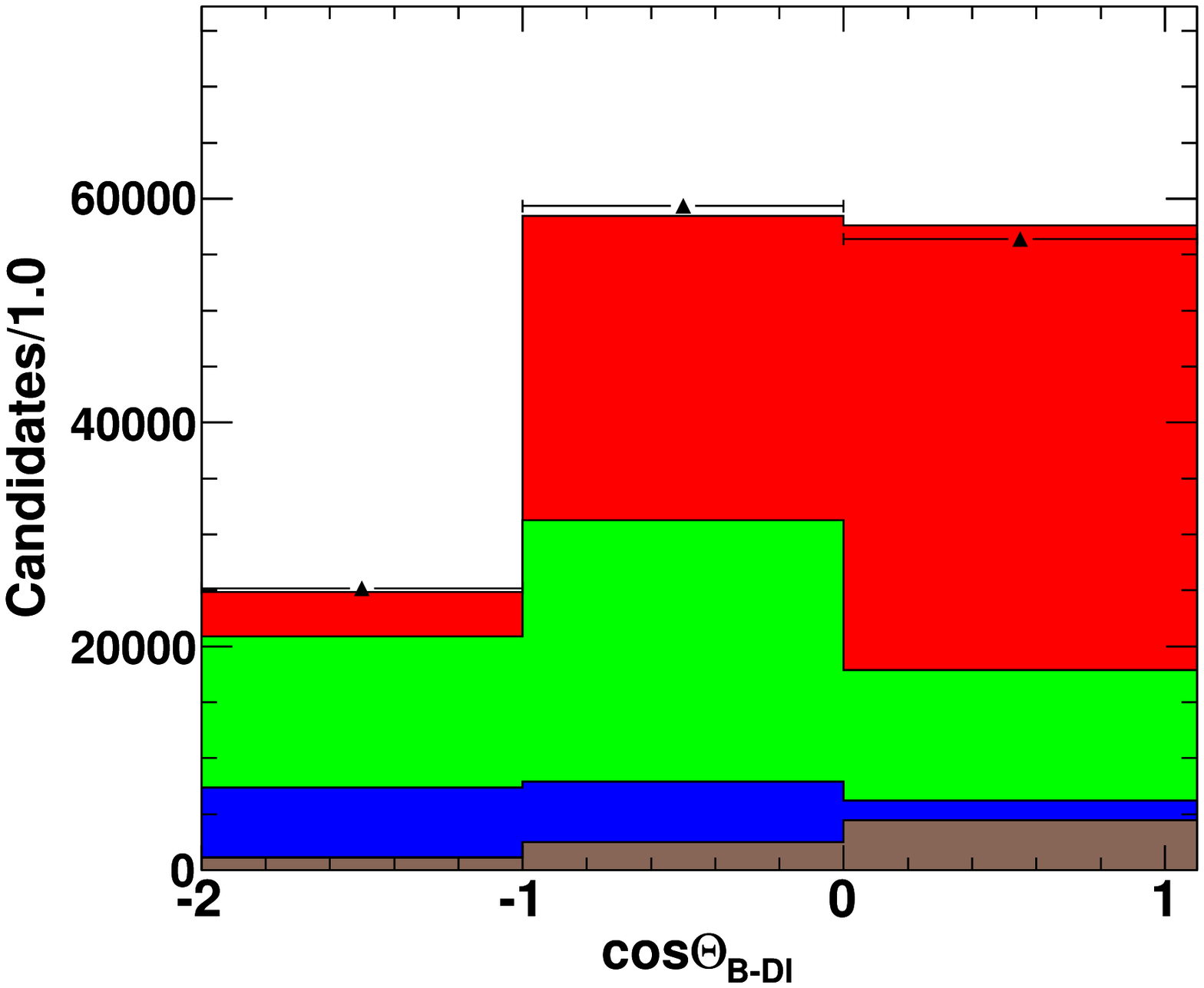}
\caption{ Projections onto individual kinematic variables of the data and the results of the fit for the $B^- \to D^0 e^-\bar{\nu}_{e}$ (left) and $\Bzb \to D^+ e^-\bar{\nu}_{e}$ (right): lepton momentum (top), $D$ momentum (middle) and $\cos\theta_{B −De}$ (bottom). The points show the data, and the histograms show the individual fit components (from top to bottom in each plot): $\Bbar \to D e^-\bar{\nu}_{e}$, $\Bbar \to D^* e^-\bar{\nu}_{e}$, $\Bbar \to D^{(*)}\pi e^-\bar{\nu}_{e}$ and other backgrounds. } 
\label{fig1}
\end{minipage}%
\begin{minipage}[t]{0.10\linewidth} 
\end{minipage}
\begin{minipage}[t]{0.50\linewidth} 
\centering 
\includegraphics[width=43mm]{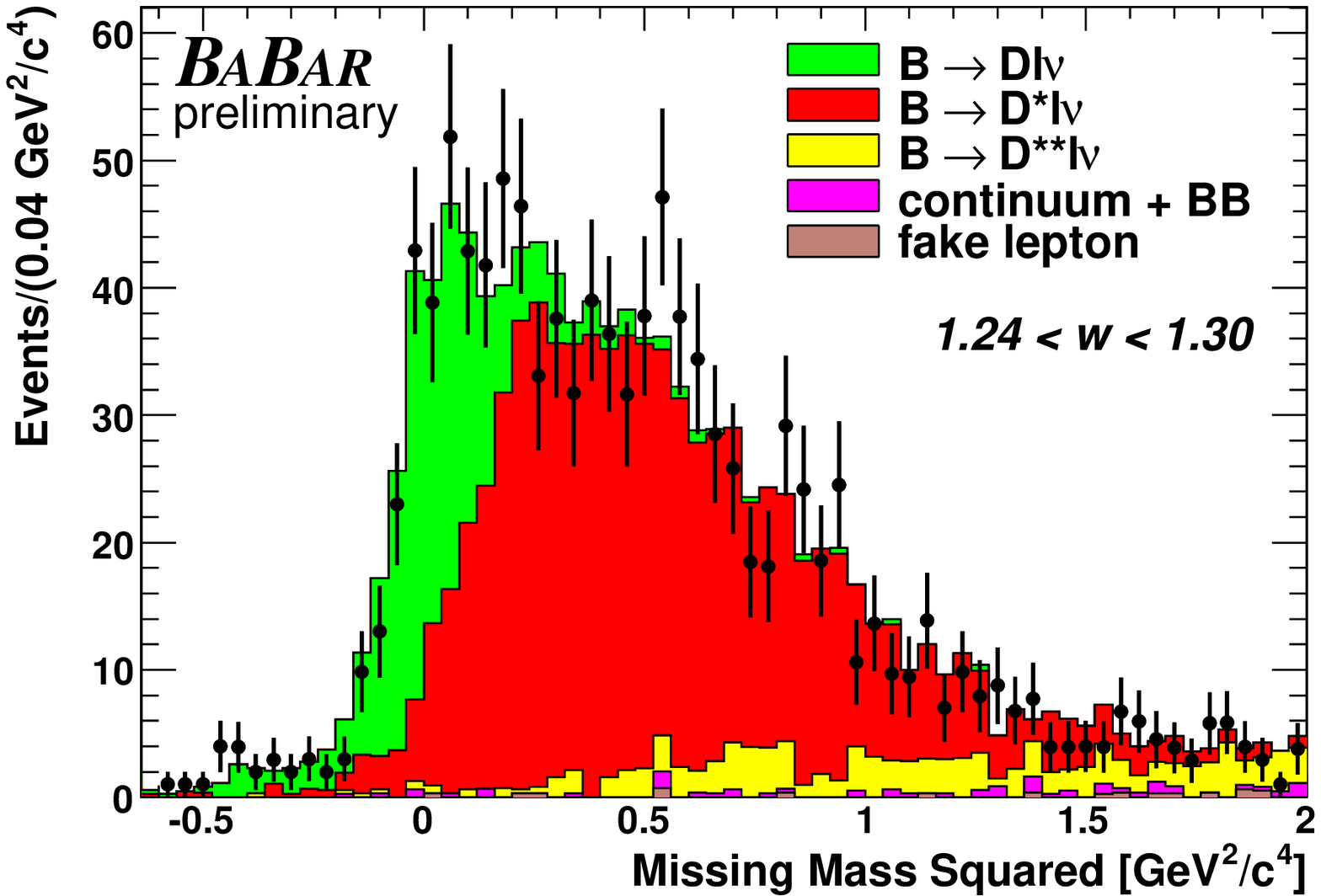}
\includegraphics[width=43mm]{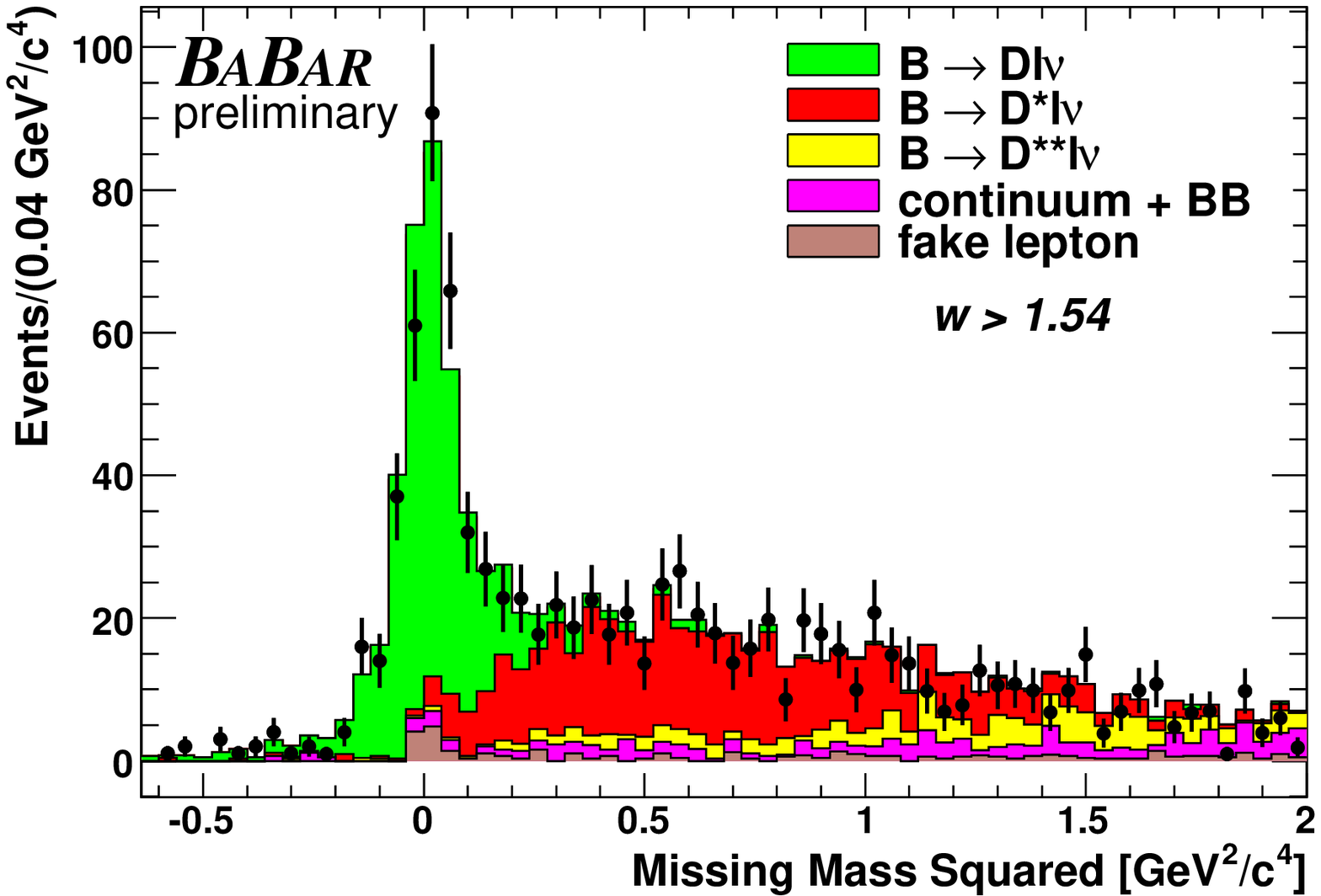}
\includegraphics[width=43mm]{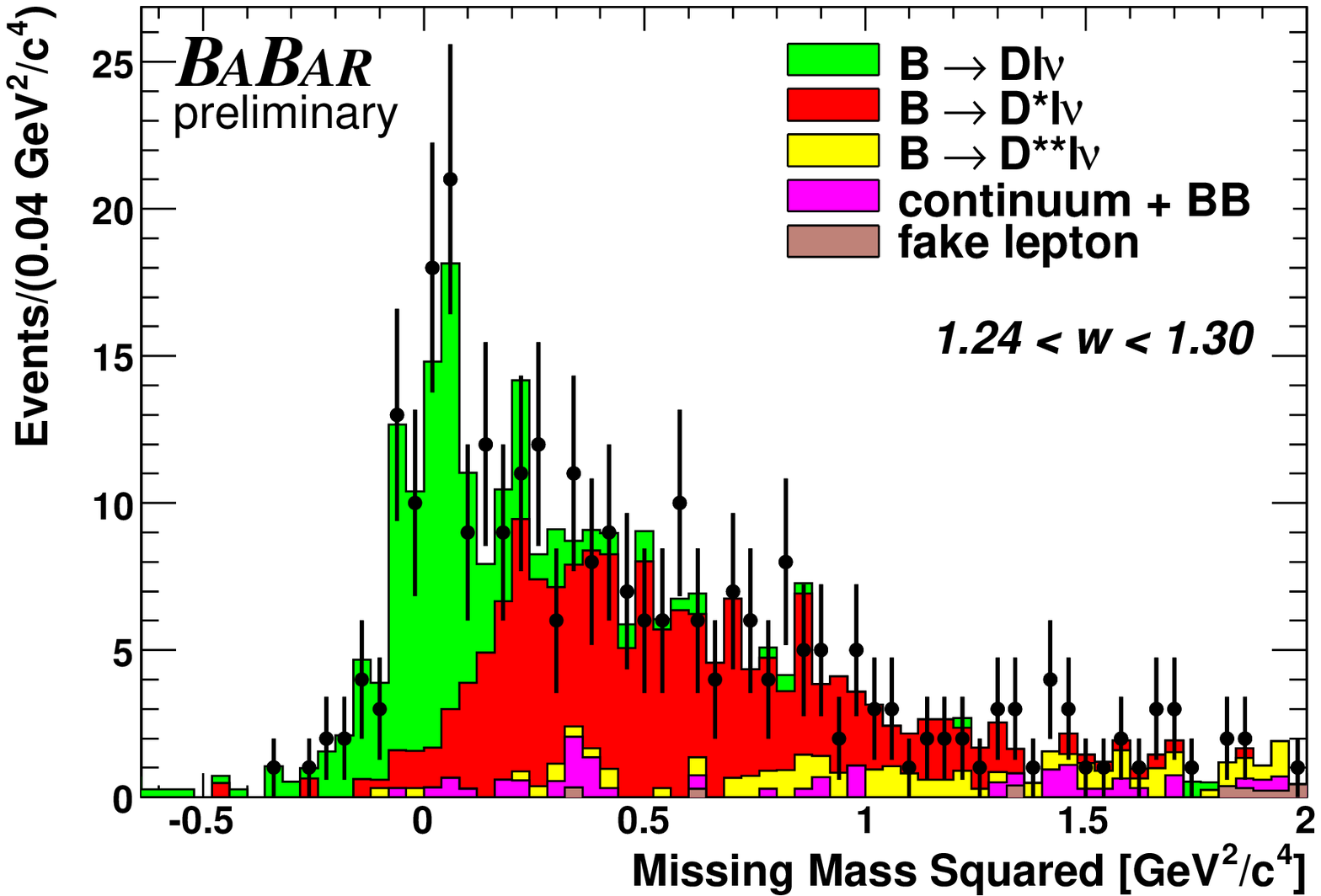}
\includegraphics[width=43mm]{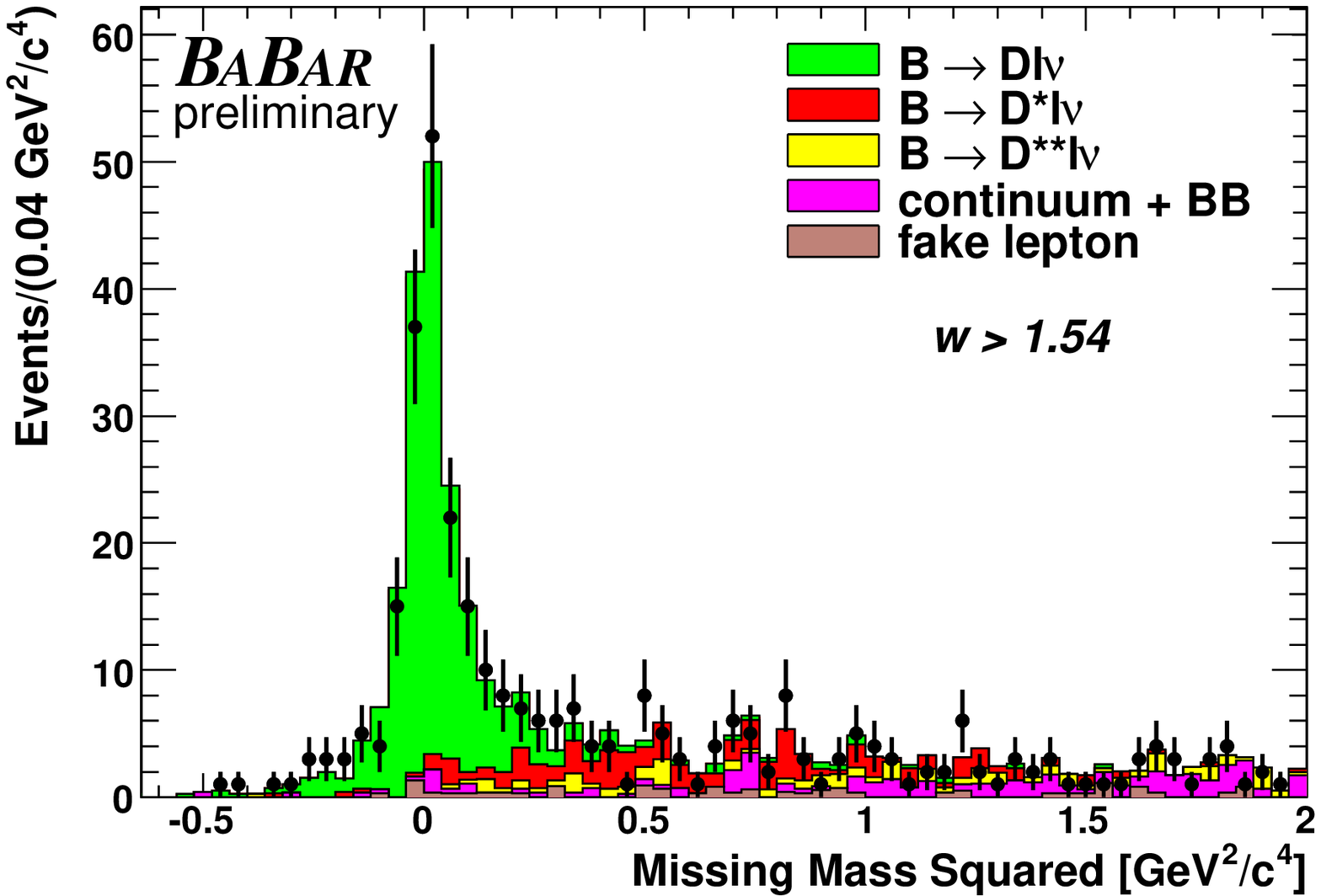}
\label{fig2}
\caption{Fit to the $m^2_\mathrm{miss}$ distribution, in two different $\om$ intervals, 
for $B^- \to D^0 \ell^- \bar{\nu}_{\ell}$ (top) and $\Bzb \to D^+ \ell^- \bar{\nu}_{\ell}$ (bottom): 
the data (points with error bars) are compared to the results of the overall
 fit (sum of the solid histograms). The PDFs for the different fit components 
are stacked and shown in different colors.}
\includegraphics[width=43mm]{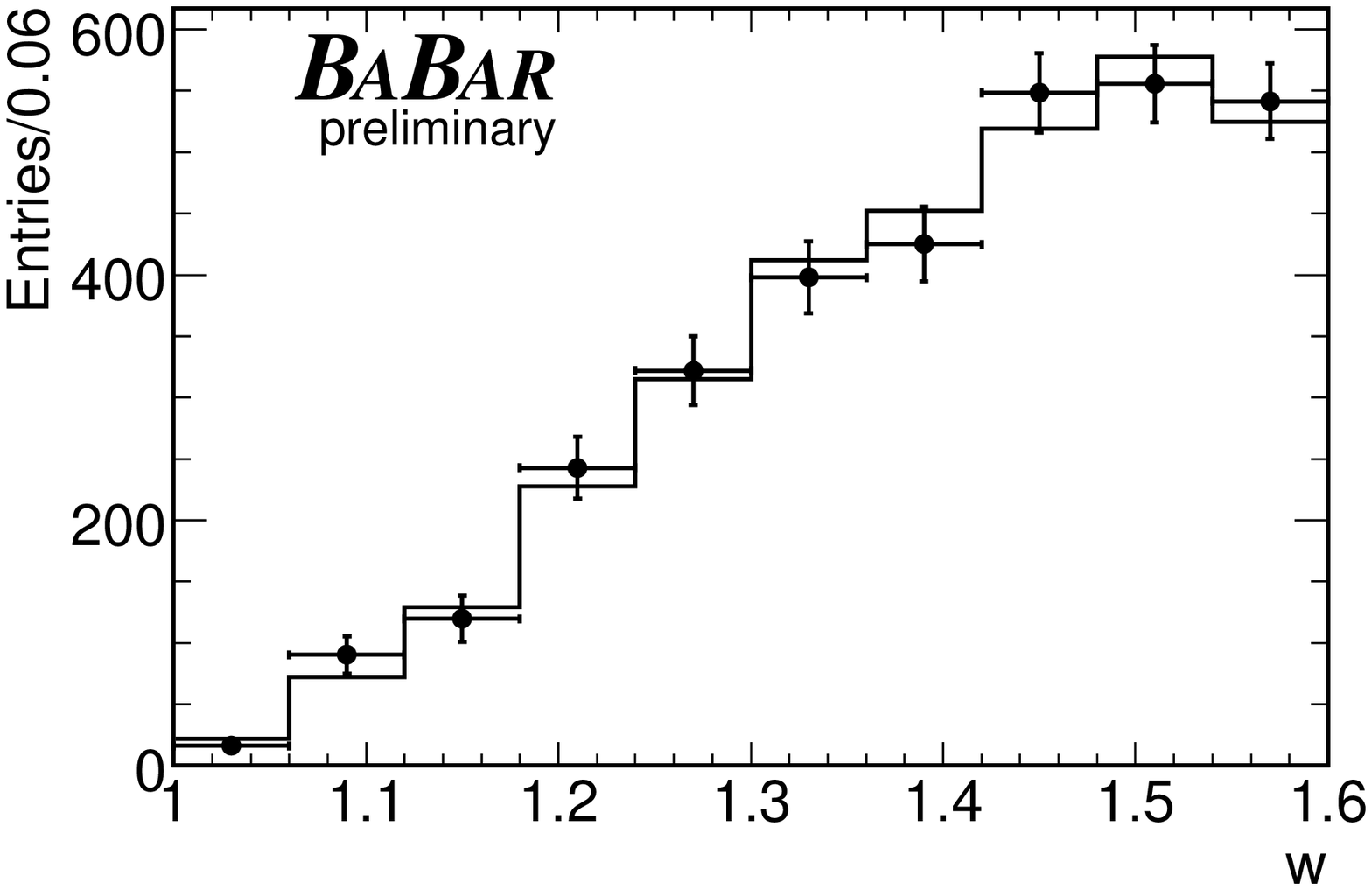}
\includegraphics[width=43mm]{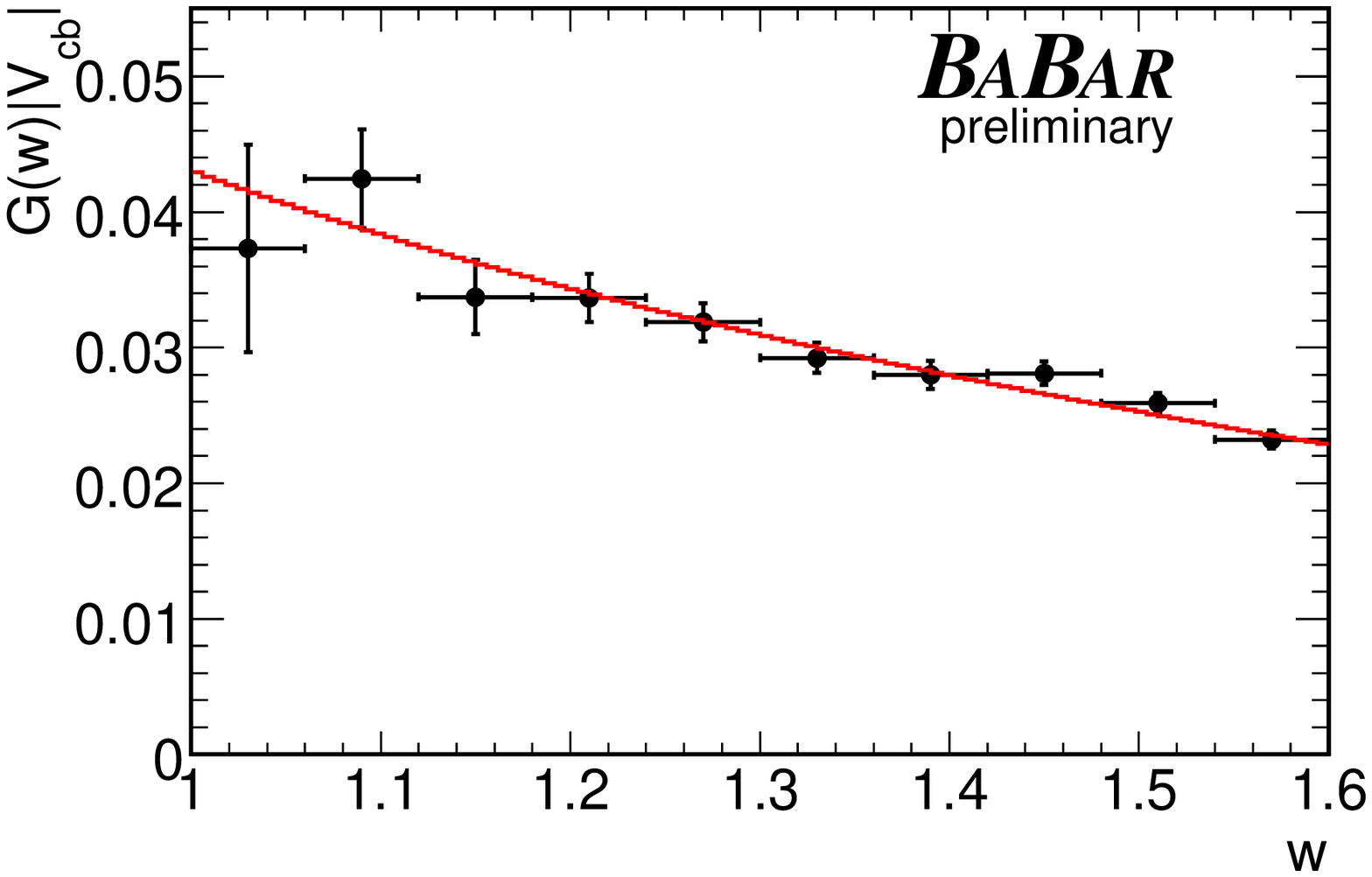}
\caption{ Left: $\om$ distribution obtained summing
the $B^- \to D^0\ell^-\bar{\nu}_{\ell}$ and $\Bzb \to D^+\ell^-\bar{\nu}_{\ell}$ yields. The data (points with error bars) are compared to the results of the fit (solid histogram). Right: ${\cal G}(w)|V_{cb}|$ distribution corrected for the 
reconstruction efficiency, with the fit result superimposed.}
\label{fig3} 
\end{minipage} 

\end{figure}

In an analysis~\cite{babarGlobal} based on  207~fb$^{-1}$ of data we reconstruct $D^0\ell^-$ and $D^+\ell^-$ pairs and 
use a global fit to their kinematic properties to determine the branching fractions and form factor parameters of the dominant semileptonic decays $\Bbar \to D^{(*)} \ell^- \bar{\nu}_{\ell}$. 
We perform a three-dimensional $\chi^2$ fit to the following variables:  $p^∗_D$, the $D$ momentum in the Center of Mass (CM) frame, $p^∗_{\ell}$, the lepton momentum in the CM frame, and $\cos \theta_{B−D\ell} = (2E^∗_{B}E∗_{D\ell}-m^2_B − m^2_{D\ell}) /(2p^∗_B p^∗_{D\ell})$, 
the cosine of the angle between the $B$ and $D\ell$ momentum vectors in the CM frame under the assumption that the $B$ decayed to $D\ell^-\bar{\nu}_{\ell}$ and the only missing particle is a neutrino. 
The energy, momentum and invariant mass corresponding to the sum of the $D$ and lepton four vectors in the CM frame are denoted $E^∗_{D\ell}$, $ p^∗_{D\ell}$, and $m_{D\ell}$, respectively. The $B$ energy and momentum are calculated from 
the CM beam energy $\sqrt{s}$ as $E^∗_B =\sqrt{s}/2$ and $p^∗_B = \sqrt{E^{*2}_B-m^2_B}$ , where $m_B$ is the $B^0$ meson mass. 
 Kinematic restrictions are imposed to reduce the contribution of backgrounds from semileptonic decays to final state 
hadronic systems more massive than $D^∗$ and from other sources of $D\ell$ combinations: we require 
$-2<\cos\theta_{B−D\ell} <1.1$,  1.2 GeV$/c < p^∗_{\ell} <2.35$ GeV$/c$ and 0.8 GeV$/c< p^∗_D <2.25$ GeV$/c$.
From the $\chi^2$ fit we measure the $\Bbar \to D \ell^-\bar{\nu}_{\ell}$ and $\Bbar \to D^* \ell^-\bar{\nu}_{\ell}$ branching fractions and 
form-factor parameters, that are then used to determine the products ${\cal G} (1)|V_{cb}|$ and ${\cal F} (1)|V_{cb}|$, where ${\cal G}(1)$ and ${\cal F}(1)$ are the form-factors at the point of zero-recoil. The $\Bbar \to D^{**} \ell^-\bar{\nu}_{\ell}$ rate is fixed in the fit to recent measurements~\cite{pdg} and the semileptonic decay rates for $B^−$ and $\Bzb$ are assumed to be equal, e.g. $\Gamma(B^- \to D^0\ell^-\bar{\nu}_{\ell}) = \Gamma(\Bzb \to D^+\ell^-\bar{\nu}_{\ell})$.  
The one-dimensional projections with the fit results for the $De$ sample are shown in Fig. \ref{fig1}.

We measure the branching fractions ${\cal B}(B^- \to D^0\ell^-\bar{\nu}_{\ell}) = (2.36 \pm 0.03_{stat.}
 \pm 0.12_{syst.})\%$ and ${\cal B}(B^- \to D^{*0} \ell^-\bar{\nu}_{\ell}) = (5.37 \pm 0.02_{stat.} \pm 0.21_{syst.})\%$ and the form factor parameters  $\rho^2_D = 1.22 \pm 0.04_{stat.} \pm 0.07_{syst.}$ ($\rho^2_{D^*} = 1.21 \pm 0.02_{stat.} \pm 0.07_{syst.}$) for $\Bbar \to D\ell^-\bar{\nu}_{\ell}$ ($\Bbar \to D^*\ell^-\bar{\nu}_{\ell}$). From these, we extract  ${\cal G} (1)|V_{cb}| = (43.8 \pm 0.8_{stat.} \pm 2.3_{syst.}) \times 10^{-3}$, and ${\cal F} (1)|V_{cb} | = (35.7 \pm 0.2_{stat.} \pm 1.2_{syst.}) \times 10^{-3}$.

\section{{\boldmath $\Bbar \to D^{(*)}(\pi)\ell^-\bar{\nu}_{\ell}$} Decays}
\label{tag1}

A measurement of the $\Bbar \to D^{(*)}(\pi)\ell^-\bar{\nu}_{\ell}$ branching fractions  has been performed~\cite{babarTagged} on 
a data sample of about 341~fb$^{-1}$. We select semileptonic $B$ decays in events containing
a fully reconstructed $B$ meson ($B_{tag}$), which allows us to constrain the kinematics, reduce the combinatorial background, 
and determine the charge and flavor of the signal $B$ meson. We reconstruct $B_{tag}$ decays of the type $\Bbar \rightarrow D Y$, where 
$Y$ represents a collection of hadrons with a total charge of $\pm 1$, composed
of $n_1\pi^{\pm}+n_2 K^{\pm}+n_3 K^0_S+n_4\pi^0$, where $n_1+n_2 \leq  5$, $n_3
\leq 2$, and $n_4 \leq 2$. Using $D^0(D^+)$ and $D^{*0}(D^{*+})$ as seeds for $B^-(\Bzb)$ decays, we reconstruct in total about 1000 
different decay chains.
The exclusive semileptonic $B$ decays are identified by the missing mass squared in the event, $m^2_{miss} = (p(\Upsilon(4S)
) -p(B_{tag}) - p(D^{(*)}(\pi)) - p(\ell))^2$, defined in terms of the particle
four-momenta in the CM frame of the reconstructed final
states. To determine the $B$ semileptonic signal yields, we perform a one-dimensional extended binned maximum likelihood fit~\cite{Barlow} to the $m^2_{miss}$ distributions. To reduce the systematic uncertainties, we measure the exclusive ${\cal B} (\Bbar \rightarrow D^{(*)}(\pi) \ell^- \bar{\nu}_{\ell})$ branching fractions relative to the inclusive semileptonic branching fraction.  
The accuracy of the branching fraction measurements for the $\Bbar \rightarrow D^{(*)}(\pi) \ell^- \bar{\nu}_{\ell}$ decays is comparable to that of the current world average~\cite{pdg}. 
By comparing the sum of the measured branching fractions for $\Bbar \to D^{(*)}(\pi) \ell^- \bar{\nu}_{\ell}$ with the inclusive $\Bbar \to X_c \ell^- \bar{\nu}_{\ell}$ branching fraction~\cite{pdg}, a $(11 \pm 4)\%$ discrepancy is observed, which 
is most likely due to $\Bbar \to D^{(*)}n \pi \ell^- \bar{\nu}_{\ell}$ decays with $n>1$. 

\section{{\boldmath $|V_{cb}|$} from {\boldmath $\Bbar \to D\ell^-\bar{\nu}_{\ell}$} Decays}

We report another measurement~\cite{babarTagged2} of the CKM matrix element $|V_{cb}|$ and the form-factor slope $\rho^2$ for  
$\Bbar \to D \ell^- \bar{\nu}_{\ell}$ decays based on  417 fb$^{-1}$ of data, using semileptonic decays in \BB\ events in which the
hadronic decay of the second $B$ meson is fully reconstructed.
The event reconstruction and the selection of the hadronic tag are similar to the analysis on the recoil~\cite{babarTagged} described in \S \ref{tag1}. 

We perform a $\chi^2$ fit to the $\om$ distribution for $\Bbar \to D\ell^-\bar{\nu}_{\ell}$ decays, where $\om$ denotes the product of the $B$ and $D$ meson four-velocities $V_B$ and $V_D$, $\om = V_B\cdot V_D=\frac{(M_{B}^2 + M_{D}^2 - q^2)}{(2M_{B} M_{D})}$, 
where $q^2 \equiv (p_{B}-p_{D})^2$, and $p_B$ and $p_D$ refer to the four-momenta of the $B$ and $D$ mesons.  Its 
lower limit, $\om=1$, corresponds to zero recoil of the $D$ meson, $i.e.$ the maximum $q^2$. 
To obtain the semileptonic $\Bbar \to D\ell^-\bar{\nu}_{\ell}$ signal yield in the different $\om$ intervals, 
we perform a one-dimensional extended binned maximum likelihood fit~\cite{Barlow} to the $m^2_\mathrm{miss}$ distributions. 
The $m^2_\mathrm{miss}$ distributions for two different $w$ intervals are compared with the results of the 
fits in Fig.~\ref{fig2}.
The comparison between the data and  the fit results for the $\om$ distribution is shown in Fig. \ref{fig3}.
We measure the branching fractions ${\cal B}(B^- \to D^0\ell^-\bar{\nu}_{\ell}) = (2.34 \pm 0.07_{stat.}
 \pm 0.07_{syst.})\%$ and the form factor parameters  $\rho^2_D = 1.20 \pm 0.09_{stat.} \pm 0.04_{syst.}$, and we extract ${\cal G}(1) |V_{cb}|= (43.0 \pm 1.9_{stat.} \pm 1.4_{syst.})\times 10^{-3}$.

\section{{\boldmath $\Bbar \to D^{**}\ell^-\bar{\nu}_{\ell}$} Decays}

In another analysis~\cite{babarTagged3} in which events are selected by fully reconstructing one of the $B$ mesons in a hadronic decay mode, we measure the branching fractions of $\Bbar \to D^{**} \ell^- \bar{\nu}_{\ell}$ decays based on 417~fb$^{-1}$ of data. 
The event reconstruction and the selection of the hadronic tag are similar to the previous analysis on the recoil. 
$D^{**}$ mesons are reconstructed in the $D^{(*)}\pi^{\pm}$
 decay modes. Semileptonic $\Bbar \rightarrow D^{**}\ell^- \bar{\nu}_{\ell}$ decays are identified by $m^2_{miss}$, and the signal yields for the $\Bbar \to D^{**}\ell^- \bar{\nu}_{\ell}$ decays are extracted through a simultaneous unbinned maximum likelihood fit to the four invariant mass difference $M(D^{(*)}\pi)-M(D^{(*)})$ distributions. 
The Probability Density Functions for the $D^{**}$ signal components are determined using Monte Carlo $\Bbar \to D^{**}\ell^- 
\bar{\nu}_{\ell}$ signal events. We rely on the Monte Carlo prediction for the shape of the combinatorial and continuum background, while its yield is estimated from data.
We observe the $\Bbar \to D^{**} \ell^- \bar{\nu}_{\ell}$ decay modes corresponding to the four $D^{**}$ states predicted by
  Heavy Quark Symmetry~\cite{hqs} with a significance greater than 6 standard deviations, including 
systematic uncertainties. We find results consistent with Ref.~\cite{babarTagged} for the sum of the different $D^{**}$ branching fractions. The rate for the $D^{**}$ narrow 
states is in good agreement with recent measurements~\cite{D0,babarUntagged}, the one for the broad states is in agreement with DELPHI~\cite{delphi2005}, but does not agree with the $D'_1$ limit of Belle~\cite{belle}. The rate for the 
broad states is found to be large.

\section{{\boldmath $\Bbar \to \Lambda^+_c X \ell^-\bar{\nu}_{\ell}$} Decays}

The $B$ decays to charmed baryons are not as well understood as the decays into
charmed mesons.  In particular, there is limited knowledge, both theoretical and experimental, about the $B$ semileptonic decays into charmed baryons.
We report the first evidence~\cite{lambdac} for $\overline{B}\to \Lambda_c^+ X e^- \overline{\nu}_{e}$
decays, where $X$ can be any particle(s) from the $B$ semileptonic decay other than the leptons and the $\Lambda_c^+$.
The $\overline{B} \to \Lambda_c^+ X e^- \overline{\nu}_{e}$ signal yield is
obtained by a one-dimensional
binned maximum likelihood fit to the $\Lambda^+_c$ invariant mass distribution. 
Monte Carlo studies show that the peaking background (events with a correctly-reconstructed $\Lambda^+_c$) comes mainly from hadronic $\overline{B}\to \Lambda^+_c X$ decays, with a fake electron from gamma conversions or $\pi^0$ Dalitz decays.
The number of peaking background events from hadronic $\overline{B}\to \Lambda^+_c X$ decays is estimated from the simulation.
We measure the relative branching fraction ${\cal B}(\overline{B}  \to \Lambda^+_c X \ell^- \overline{\nu}_{\ell})/{\cal B}(\overline{B} \to
  \Lambda^+_c/\overline{\Lambda}^-_c X)=    (3.2\pm0.9_{\rm stat.}\pm0.9_{\rm syst.})\%$, corresponding to a significance of the signal, taking into account the systematic uncertainty, of 4.9 standard deviations.

\section{Conclusions}

We have reported several new measurements of exclusive $\Bbar \to X_c \ell^- \bar{\nu}_{\ell}$ decays and the $|V_{cb}|$ CKM matrix element obtained by the \babar\ experiment. Including these new \babar\ measurements, the HFAG~\cite{excl} obtains the averages ${\cal G}(1)|V_{cb}| = 42.4 \pm 1.6$ and ${\cal F}(1)|V_{cb}| = 35.41 \pm 0.52$. The accuracy on the $|V_{cb}|$ extraction from $\Bbar \to D \ell^-\bar{\nu}_{\ell}$ decays is highly improved by the recent \babar\ results~\cite{pdg}.  Using the form-factor normalizations~\cite{lat} ${\cal G}(1)=1.074 \pm 0.018 \pm 0.016$ and ${\cal F}(1)=0.921 \pm 0.013 \pm 0.020$, we obtain $|V_{cb}|=39.2 \pm 1.5 \pm 0.9$ and $|V_{cb}| = 38.2 \pm 0.6 \pm 1.0$, for $\Bbar \to D \ell^- \bar{\nu}_{\ell}$ and $\Bbar \to D^* \ell^- \bar{\nu}_{\ell}$, respectively, and the third error is due to the form-factor normalization. Comparing with the inclusive $|V_{cb}|$ determination~\cite{excl}, $|V_{cb}| = 41.67 \pm 0.73$, we observe that some tension is still present, more significant for the $\Bbar \to D^* \ell^- \bar{\nu}_{\ell}$ determination than for the $\Bbar \to D \ell^- \bar{\nu}_{\ell}$ one.  It should also be noted that quenched lattice deteminations~\cite{nazario} of ${\cal F}(1)$ and ${\cal G}(1)$  tend to favor higher $|V_{cb}|$ values.  

The accuracy on the exclusive $\Bbar \to X_c \ell^- \bar{\nu}_{\ell}$ branching fraction measurement has reached the 3-4\% level for $\Bbar \to D^{(*)} \ell^- \bar{\nu}_{\ell}$ and the 10\% level for $\Bbar \to D^{**} \ell^- \bar{\nu}_{\ell}$. A 10\% difference between the sum of the $\Bbar \to D^{(*,**)} \ell^- \bar{\nu}_{\ell}$ rates and the inclusive rate is observed: in order to resolve this puzzle, additional measurements at the $B$-factories are required.


\end{document}